\documentstyle[twocolumn,prl,aps]{revtex}
\begin{document}
\draft
\title{ Dynamic Domains above the Ferromagnetic Resonance
Instability}
\author{T. Plefka}
\address{ Theoretische Festk\"orperphysik, Technische Hochschule

Darmstadt, 
D 64289  Darmstadt, Germany}
\date{\today}
\maketitle
\begin{abstract}
A simplified model is introduced and analysed to show, that  for the
Landau-Lifshitz  equation stable, steady state solutions of

domain type exist
in ferromagnetic systems, strongly driven by external transverse
fields. These  dynamic domain states are able to describe the
drastic
reduction of the power absorption found experimentally above the
instability
of the homogeneous magnetisation. The excitations of the domain
states are
presented and the relevance of the model for real ferromagnets is
discussed.
The analogy of this driven
dissipative system to equilibrium systems of coexisting
phases is worked out.
\end{abstract}

Ferromagnetic systems strongly driven by external transverse
magnetic
fields are under  investigation since more than four decades (for
the

early work see \cite{da63}). Based on a nonlinear spin-wave
expansion around the homogeneous ground state Suhl \cite{su57}
showed
that certain modes  become unstable for increasing amplitude of the

pump field.
For a special case Suhl  proposed  an off threshold
steady

state leading to
an explanation for the saturation of the power
absorption
even far above threshold.

Detailed measurements of Gibson and Jeffries \cite{gi84} and
further
experiments \cite{ex} exhibit beyond threshold many
of the  phenomena predicted by the general theory of  non-linear
dynamics. Based on \cite{su57}  truncated
spin-wave mode models \cite{moden}
have been published, which with some success describe the
experimental
findings \cite{com1}. The arguments leading to these
models are usually based on {\it ad hoc} assumptions which can at
best be
justified \cite{Su86} in a small vicinity  near threshold or for the

special case of films (see e.g.  McMichael and
Wigen \cite{moden}). Thus at least for bulk materials
these approaches are not  satisfactory.

Very recently Cross and Hohenberg
\cite{cr93}  questioned the off threshold approach
of Suhl and raised doubts against the proposal of
Anderson \cite{an81} that the ferromagnetic
resonance instability is a characteristic example for the general
problem
how the concepts of equilibrium phase transitions can be extended
 to driven dissipative systems.

In this letter I will focus  on regions of the control parameters
above
threshold
where the  power absorption is time independent. Such
a behaviour  is experimentally well
established often in coexistence with other time dependent
states. At low temperatures
exclusively time independent states have
been observed by Wiese {\it et al.} \cite{ex} for spherical
samples of yttrium iron garnet.

I will point out that above threshold stable steady state
solutions of the equation of motion exist with domain character.
These solutions are able to explain the saturation effect in
the absorption and are therefore an alternative to the off
threshold
approach of \cite{su57}. Dynamic domain states have already been
found in a computer simulation by Elmer \cite{el88}. Static domain
states are well established and although the theory \cite{dom} of
ferromagnetic domains  has been worked out to a
high extent, dynamic problems have only been treated in linear
response  approximation \cite{domlin}.

At a mesoscopic scale the dynamics  is governed by the
Landau-Lifshitz
equation as recently demonstrated by microscopic investigations
 \cite{pl90,ga91}. In the frame, rotating with \bbox{ \omega},
 this equation of motion takes the form
\begin{equation} \label{1}
{\bf \dot{m}}\,=-{\bf m\times} \biglb( {\bf H}^{{\rm eff}}\,-
 \bbox{ \omega}
\biglb)\,- \Gamma \,{\bf m\times \bigl( m \times H }^{{\rm eff}}
 {\bf  \bigl)}\,.
\end{equation}
$ {\bf m}({\bf r},t) $ and ${\bf H}^{{\rm eff}}$ are the local
magnetisation and the effective field in the
rotating frame, being related to the quantities
$ {\bf m}_{{\rm lab}} $ and  ${\bf H}^{{\rm eff}}_{{\rm lab}} $ in
the laboratory frame by
$ {\bf m}_{{\rm lab}}\,=\, \exp (\,t {\bbox {\omega
 \times}}\,)\,\,{\bf m} $ and by
$ {\bf H}^{{\rm eff}}_{{\rm lab}}\,=\, \exp (\,t {\bbox {\omega
\times}}\,)
\,{\bf H}^{{\rm eff}} $,
respectively. $ \Gamma $ represents the Landau-Lifshitz damping rate
and

in reduced units the gyromagnetic ratio and the magnitude of the

magnetisation  $m\,=\, |{\bf m}|$ are equal to 1. For the effective

field it is assumed
\begin{equation} \label{2}
{\bf H}^{{\rm eff}}\,=\, {\bf h}_\parallel \, + \,{\bf h}_\perp \,-

\, \bbox{\overline{m}} \;,
\end{equation}
where ${\bf h}_\parallel $ is the dc external field and where

 ${\bf h}_\perp $
is the amplitude of the circular rf  field,  which are parallel and

transverse to driving frequency $ \bbox{ \omega} $ respectively. The

term $ \bbox{\overline{m}}\,= \,V^{-1}\int{\bf m}\,{\rm d}V $
represents,

again in reduced units, the demagnetisation field of a sphere of
volume $ V $. 
No explicit time dependence arises in the rotating
frame
being the main advantage of this frame.

Eqs.(\ref 1) and (\ref 2) describe a model, which is certainly
oversimplified for realistic ferromagnetic materials as the non
local
contribution to the exchange field, parts of the dipolar
field and anisotropy fields are omitted. This  model,
however, is  believed to describe the basic features,
even in  regions far above  threshold, and this model
can be discussed analytically.

First the special case of a pure {\it static field} is considered
($ \bbox{ \omega}= 0,\;{\bf h}_\parallel \, +{\bf h}_\perp
\rightarrow {\bf h}_0 $).  For this case the energy
$ E\,= \, \bbox{\overline{m}}^2/2\;-\, \bbox{\overline{m}} {\bf h}_0
$

is a Lyapunov functional of  $ {\bf m}({\bf r},t) $ \cite{com},
from which it is immediately concluded that the equation of motion

has stable
fix points  satisfying  $ \bbox{\overline{m}}={\bf h}_0 $

for $  h_0<1 $  and satisfying  $ \bbox{\overline{m}}={\bf h}_0/
h_0$

for $  h_0>1 $, respectively. As $ m=1$ holds this result implies
that

for $  h_0<1 $ these solutions are of domain type and highly
degenerate,

while for  $  h_0>1 $ the magnetisation is homogeneous. Thus the

simplified model exhibits domains for static fields being a basic
 requirement on such a model. Note that  it is the internal field
$  {\bf H}^{\rm int} ={\bf h}_0-\bbox{\overline{m}}$
which is identified as the ordering field conjugate
the order parameter $ \bbox{\overline{m}}$ in the sense of phase

transitions. Then the transition occurs at $ H^{\rm int}=0 $ with
all
 the characteristic features of a first order phase transition, as
e.g.
 coexisting phases.

Coming to the central point to analyse eq.(\ref{1}) in the
general case for  fix points
the volume of the sample is partitioned in partial volumes
$ V_i = n_i V $ in which $ {\bf m}({\bf r}) $    has the
values $ {\bf m}_i $. As the index $i $ may be read
as $ {\bf r} $, this ansatz does not imply any restriction and leads
to
\begin{equation} \label{3}
{\bf  H}_1 \,+ \Gamma {\bf m}_i{\bf \times H }_2 \, =
\, \alpha_i{\bf m}_i
\end{equation}
where $ \alpha_i $ and where
\begin{equation} \label{4}
{\bf  H}_2={\bf  H}_1+ \bbox{ \omega} ={\bf h}_\parallel
+{\bf h}_\perp-\sum n_i {\bf  m}_i
\end{equation}
was introduced. Formally solving eq.(\ref{3}) for $ {\bf  m}_i $
results in
\begin{equation} \label{5}
{\bf m}_i =  \frac{ \alpha_i {\bf H}_1 + \alpha_i ^{-1} \Gamma^2
\bigl({\bf H}_1 {\bf H}_2 \bigr){\bf H}_2\,+ \Gamma{\bf  H}_1{\bf

\times H}_2}
{\Gamma^2 H^2_2 \, + \, \alpha_i^2}\,.
\end{equation}
Eq.(\ref{3}) multiplied by ${\bf m}_i$ gives $ \alpha_i  ={\bf m}_i
{\bf H}_1 $ and using in addition eq.(\ref{5}), leads to a

biquadratic equation
for  $ \alpha_i $. This equation has two real solutions
 $ \alpha_\pm = \mp \alpha $   where $ \alpha(>0) $ is determined by

\begin{eqnarray}\label{6}
2 \alpha^2 &  = &  H_1^2-\Gamma^2 H_2^2 \nonumber \\ &+&
\Bigl[ \bigl( H_1^2-\Gamma^2 H_2^2 \bigr)^2+
4 \Gamma^2 \bigl({\bf H}_1{\bf H}_2 \bigr)^2 \bigr]^{1/2}.
\end{eqnarray}
As $i $ can take only the values $ + $ and $ - $,  there are only
two
possible values $ {\bf m}_+$ and $ {\bf m}_-$  for local
magnetisation being realised in the, in general
disconnected, partial volumes $V_+$ and $ V_-$, respectively.
Reading the index  $ i $  as $ {\bf r} $, it can be concluded,
that apart from these domain states no other fix points
are possible.  Note that eqs.(\ref{4})-(\ref{6}),
together with $ n_+ +n_-=1 $, determine $ {\bf m}_i $
for given ${\bf h}_\parallel ,\,{\bf h}_\perp ,\,  \bbox{ \omega},
\Gamma $ and for given $ n_+$. As these eqs. are implicit,
a discussion is needed for which values of the parameters solutions

do exist. Focusing  on stable solutions I first will put forward

the stability analysis
which simplifies this discussion.

For this analysis
$ {\bf m} ({\bf r},t)= {\bf m}_i +{\bf d}_i ({\bf r},t) $
is introduced in $ V_i $ and eq.(\ref{1}) is linearised
in the deviations $ {\bf d}_i  $. With the transformation
\begin{eqnarray}
{\bf \widetilde{d}}_i & = & {\bf d}_i - n_i^{-1} {\bf
\overline{d}}_i

\nonumber \\
{\bf \overline{d}}_i & = & V^{-1}\int_{V_i} {\bf d}_i ({\bf
r},t)\,{\rm d}V
\label{8}
\end{eqnarray}
the linear system partly decouples resulting in
\begin{eqnarray}\label{9}
{\bf \dot{\widetilde{d}}}_i & = & \alpha_i {\bf m}_i
{\bf \times \widetilde{d}}_i - \Gamma \bigl( {\bf H}_2  {\bf m}_i
\bigr)
{\bf \widetilde{d}}_i \\
{\bf \dot{\overline{d}}}_\pm  & = & (\alpha_\pm + n_\pm )
{\bf m}_\pm{\bf \times \overline{d}}_\pm + n_\pm {\bf m}_\pm {\bf
\times

\overline{d}}_\mp \nonumber \\
& {} & -\Gamma \bigl({\bf H}_2  {\bf m}_\pm  + n_\pm \bigr)
{\bf \overline{d}}_\pm \nonumber \\
& {} & + \, \Gamma n_\pm{\bf m}_\pm{\bf \times}\bigl(
{\bf m}_\pm {\bf \times}  {\bf \overline{d}}_\mp
 \bigr) \label{10}
\end{eqnarray}
where $ {\bf m}_i {\bf d}_i=0 $ was used being a consequence of $
m=1 $.

According to eq.(\ref{9})  the damping of  excitations
$ {\bf \widetilde{d}}_i $ is given by $  \Gamma ({\bf H}_2  {\bf
m}_i) $
which is rewritten with eq.(\ref{5}) as $  \Gamma ({\bf H}_1{\bf
H}_2)/
\alpha_i $. Recalling that   $ \alpha_i  $ can take the values $ \pm
\alpha $
it is immediately concluded that stable fix point solutions are only

possible for two special situations. This is the
{\it homogeneous state } characterised by $ ({\bf H}_1{\bf H}_2 )
\neq  0$ and $ n_+ =1 $ , and this are the {\it domain states}
characterised
by $ ({\bf H}_1{\bf H}_2)  = 0  $  and allowing   solutions with $
n_+ <1 $.
Note that the condition  $({\bf H}_1{\bf H}_2)  = 0  $
represents the extension of $ H^{int}=0 $ from the static to
the dynamic case.

All the essentials of the present approach have been deduced.
For the remaining  point, to work out the  results explicitly,
I change notations for reasons of more transparency. The homogeneous
state,
which arises up to here as a limit of the domain states, will be
denoted

by the
index "hom ", reserving "$ \pm $ "for  true  domain states.

For {\it the homogeneous state}, setting $ \hat{\alpha}_{{\rm hom}}=

\alpha_{{\rm hom}} +1 $ and
introducing the fields
\begin{equation} \label{11}
{\bf  h}_2={\bf  h}_1+ \bbox{ \omega} ={\bf h}_\parallel +{\bf
h}_\perp \;,
\end{equation}
eq.(\ref{3}) is rewritten  as
$ {\bf  h}_1 \,+ \Gamma {\bf m}_{{\rm hom}}{\bf \times h }_2 \, =
\,
\hat{\alpha}_{{\rm hom}} {\bf m}_{{\rm hom}}$ . Recalling that the
steps
from eq.(\ref{3}) leading to eqs.(\ref{5}),(\ref{6}) are purely

algebraic, the solution
for $ {\bf m}_{{\rm hom}} $ is given by these eqs. with the
replacements
$ {\bf  H}_1 \rightarrow {\bf  h}_1,\,{\bf  H}_1 \rightarrow {\bf
h}_2 $
and $ \alpha_i \rightarrow \hat{\alpha}_{{\rm hom}} $. Thus
 values of $ {\bf m}_{{\rm hom}}$
 and of  $ \hat{\alpha}_{{\rm hom}} $  have explicitly been found.

Assuming an $ \exp(\lambda t ) $ time dependence, the frequencies
and the dampings of the  excitations are calculated from
eqs.(\ref{9}),
(\ref{10}) in terms of  $ {\bf  h}_1 $ and $ {\bf  h}_2 $ to
\begin{eqnarray}\label{12}
\widetilde{\lambda}_{{\rm hom}} & = & \pm  i (\hat{
\alpha}_{{\rm hom}} -1) -  \Gamma \Bigl( \frac{ |{\bf h}_1 {\bf
h}_2|}
{\hat{\alpha}_{{\rm hom}}}-1 \Bigr)\\
\label{13}
\overline{\lambda}_{{\rm hom}} & = & \pm  i \hat{
\alpha}_{{\rm hom}}  -   \frac{ \Gamma }
{\hat{\alpha}_{{\rm hom}}}|{\bf h}_1 {\bf h}_2|\,.
\end{eqnarray}
$ \Gamma $ and $  \hat{\alpha}_{{\rm hom}}  $ are
positive quantities by definition. Thus the collective

excitations $ {\bf  \overline{d}}_
{{\rm hom}} $ are always damped, whereas the highly degenerate
modes
$ {\bf  \widetilde{d}}_ {{\rm hom}}({\bf r},t) $ are damped only for

$  |{\bf h}_1 {\bf h}_2| > \hat{\alpha}_{{\rm hom}} $ . This
condition can be shown to be equivalent to $ f>1 $ where
\begin{equation} \label{14}
f({\bf  h}_1,{\bf  h}_2) = \frac{({\bf  h}_1
{\bf  h}_2)^2 + \Gamma^2 h_2^2}{h_1^2 + \Gamma^2}.
\end{equation}
Thus the homogeneous state is stable for all values of $ {\bf h}_1
$ and  $ {\bf h}_2 $ for which $ f>1 $ holds. At $ f=1 $ this
state becomes unstable via  Hopf bifurcation. This
instability is a very rudimental form of the Suhl

instabilities \cite{su57}.

Next I consider the {\it domain states}. Eq.(\ref{5}) simplifies
with
$ ( {\bf H}_1{\bf H}_2)=0 $  to
\begin{equation} \label{15}
{\bf m}_\pm = \mp \sqrt{1-m_3^2}\, {\bf e}_1 +m_3{\bf e}_3 \; ;
\quad m_3= \Gamma  H_2/H_1
\end{equation}
where the orthonormal system   $ {\bf e}_1={\bf H}_1/H_1\,,
{\bf e}_2={\bf H}_2/H_2 $ and $ {\bf e}_3={\bf e}_1{\bf
\times e}_2 $ was introduced. For the bulk
magnetisation in a domain state
\begin{equation} \label{16}
{\bf \overline{m}}_{{\rm dom}}= -(n_+-n_-) \sqrt{1-m_3^2}\,

{\bf e}_1 + m_3 {\bf e}_3
\end{equation}
is found. Employing  the eqs.(\ref{4}),(\ref{11}),(\ref{15}) and
(\ref{16})
\begin{eqnarray}\label{17}
\overline{m}_{{\rm dom}}^2&  = &f({\bf h}_1, {\bf h}_2)\\
\label{18}
4n_ + n_- & = &  \frac{1-  \overline{m}_{{\rm dom}}^2} {1-m_3^2}\\
\label{19}
 H_1^2& = & \frac{\bigl( {\bf h}_1({\bf h}_1-{\bf h}_2)\bigr)
^2+ \Gamma^2({\bf h}_1-{\bf h}_2)^2}{h_1^2 + \Gamma^2}
\end{eqnarray}
and
\begin{equation}
\label{20} H_2^2 =   h_2^2-\overline{m}_{{\rm dom}}^2
\end{equation}
is obtained. As
$  \overline{m}_{{\rm dom}} $ can not be greater  than 1,
eq.(\ref{17}) implies that {\it domain states  exist only
for} $ f<1 $. Note that $ n_+ $  and all the vector components
 of eq.(\ref{15}) and of eq.(\ref{16}) have been expressed in terms

of $ {\bf h}_1$ and $ {\bf h}_2  $
or, according to eq.(\ref{11}), in terms of the external parameter

$  h_\parallel ,\, h_\perp $ and $ \omega $. By a
linear transformation  the vectors
${\bf m}_\pm $ and $ {\bf \overline{m}}_{{\rm dom}}$
may now be written in the frame spanned by ${\bf h}_\parallel ,\,
{\bf h}_\perp $ and by $ {\bf h}_\parallel {\bf \times h}_\perp $.
These explicit results are rather lengthy and will be published
elsewhere. Here only some characteristic properties are presented.
According to eq.(\ref{15})
$ {\bf m}_+{\bf m}_- = -1 + 2m_3^2 $ holds, which implies an
angle  smaller than $ \pi $ between the domain magnetisations.
In resonance ($ \omega =  h_\parallel $) this angle
is nearly $ \pi $, as in this case $ m_3
\rightarrow  h_\perp $, assuming in addition the typical parameter
values
$ \Gamma \ll 1$ and $ h_\perp \ll 1 $. The components of
$ {\bf m}_\pm $ parallel to the dc field are
calculated to $ \pm H_1 \omega^{-1} \bigl( 1 -
m_3^2 \bigr)^{1/2} $,
which reduces in the resonance case again with the typical
 values to $ \pm \Gamma ( \Gamma^2 + h_\perp^2)^{-1/2}$.
These results imply a strong deviation  of $ {\bf m}_-$
from the positive  $ {\bf h}_\parallel $ axis. Such  deviations

can not be 
described by the spin-wave expansions  \cite{moden}

around this axis.

Again with the $ \exp(\lambda t) $  ansatz the excitations of these

domain states are calculated from eq.(\ref{9}) to
\begin{equation} \label{21}
\widetilde{\lambda}_+ = -\widetilde{\lambda}_ - = \pm  i
\alpha_{{\rm dom}}\; ;\quad  \alpha_{{\rm dom}}   = \sqrt{H_1^2-
\Gamma^2 H_2^2},
\end{equation}
and from eq.(\ref{10}) in the weak damping limit ($ \Gamma \ll 1$)
to
\begin{equation}\label{23}
\overline{\lambda}_{1/3}  =  \overline{\lambda}_{2/4}^{\,\ast}
= i \sqrt{A \pm B}-\Gamma \bigl(\frac{1}{2}
\pm \frac{C}{B} \bigr)
\end{equation}
where
\begin{eqnarray}
A & = & \frac{1}{4} + n_+  n_-(1-2m_3^2)+ \bigl[ \alpha_{{\rm
dom}}+
\frac{n_+- n_-}{2}\bigr]^2 \nonumber \\
B^2 &  = & n_+  n_-m_3^4 \nonumber \\
&+& \bigl[1-4 n_+ n_-(1-m_3^2)^2 \bigr]
\bigl[\alpha_{{\rm dom}}+ \frac{n_+- n_-}{2}\bigr]^2  \nonumber \\
C & = & n_+  n_-m_3^4 + \frac{n_+- n_-}{2}
\bigl[ \alpha_{{\rm dom}}+ \frac{n_+- n_-}{2}\bigr].\nonumber
\end{eqnarray}
According to eq.(\ref{21}) the highly degenerate modes
$ {\bf \widetilde{d}}_\pm({\bf r},t) $ are marginal stable.
{}From eq.(\ref{23}) it can be shown that the coupled collective
modes $ {\bf \overline{d}}_+ $ and $ {\bf \overline{d}}_- $
are damped everywhere. Further investigations of other limiting
cases
than $ \Gamma \ll 1$ lead to the same conclusion.

Therefore it is concluded that {\it stable domain states do exist
 above the ferromagnetic resonance instability} ($f < 1$), which

is the main
result of this work.   These
are the only possible stable fix points of eq.(\ref{1}) apart
from the homogeneous state which is stable below the instability
 ($f >1$). The critical threshold value of
the pump field $h^{\rm crit}$ is in general implicitly determined by

$f({\bf h}_1, {\bf h}_2) =1$, but reduces in resonance
($ \omega =  h_\parallel $) and for $ \Gamma \ll 1$
to
\begin{equation} \label{26}
h^{\rm crit}= \Gamma (h_\parallel ^2-1)^{1/2}\;.
\end{equation}
In the laboratory frame the domain states are limit cycles.
The  components of  $ {\bf m}_\pm $ transverse to $ {\bf
h}_\parallel $
are precessing with the frequency $ \bbox{\omega} $
and with a fixed phase difference, whereas the parallel components

are time independent.

The results will be applied to calculate the {\it power
absorption},
which is defined as $ P = - \bbox{ \omega} {\bf \times h}_\perp

\bbox{\overline{m}} $, and may be rewritten with
eq.(\ref{4}) as  $ P = { \bf H}_1 {\bf \times H}_2
\bbox{\overline{m}} $ and  is found to be
\begin{eqnarray}\nonumber
 P_{{\rm hom}}& = &  \frac{ 2 \Gamma \bigl[ h_1^2 h_2^2-
({\bf h}_1 {\bf h}_2)^2 \bigr]}{ h_1^2 + \Gamma^2 h_2^2+
 \bigl[\bigl(   h_1^2-\Gamma^2 h_2^2 \bigr)^2\,+
4 \Gamma^2 \bigl({\bf h}_1{\bf h}_2 \bigr)^2 \bigr]^{1/2}}\\
\label{24} P_{{\rm dom}}& = & \Gamma \frac{ h_1^2 h_2^2-
({\bf h}_1 {\bf h}_2)^2}{h_1^2 + \Gamma^2}
\end{eqnarray}
for the homogeneous and the domain states, respectively.
In resonance with $ \Gamma \ll 1$
and $ h_\perp \ll 1 $ eq.(\ref{24})  reduces to
\begin{equation} \label{25}
P_{{\rm hom}} =  \Gamma ^{-1} h_\perp^2 \,;
\;P_{{\rm dom}} =  \Gamma h_\parallel^2 h_\perp^2
( \Gamma^2 + h_\perp^2  )^{-1},
\end{equation}
where the homogeneous case applies for $  h_\perp <h^{\rm crit} $
and the domain case  for $  h_\perp >h^{\rm crit} $.
Eq.(\ref{24}), in the general case, and eq.(\ref{25}),

in the resonance case,
describes a drastic reduction of the power absorption, which is
found in
all measurements \cite{da63,ex} above threshold. The results for the

absorption differ from \cite{su57}, but both approaches lead to a
{\it saturation  in the high power limit} $ h_\perp \gg \Gamma $.

Many aspects of the presented approach are
analogue to a thermodynamic system with coexisting phases
and represent a natural extension of the static case which is
such a equilibrium system: The asymptotic spatial structure
of the dynamic evolution, a microscopic property, strongly
depends on the special initial conditions  and is not
described by such an approach.
Macroscopic variables, however, like  $ \bbox{\overline{m}} $,
${\bf m}_\pm $ , $ n_{\pm} $ and $ P $, are independent
of the initial conditions and depend only
on the external parameter, and are continuos
quantities with discontinuous derivatives as function of
these parameters $ h_\parallel , h_\perp , \omega  $.

Based on these general features
and employing again the analogy to the statics it is expected
that dynamic domain states are also relevant to the real systems.
A further argument for this statement results from the
fact, that the domain states are an effect of  demagnetisation
field
describing  the long-ranged  dipole interaction, recalling in
addition
that only local and short-ranged contributions to $ {\bf H}^{{\rm
eff}} $
have to be added for realistic models.  Finally the
results of Elmer \cite{el88}  demonstrate
the importance of the domain states, as in this simulation
a more realistic model was employed.

The modifications in realistic models are expected
to be moderate for macroscopic properties and far
above the instability, but are expected to be more drastic for local

properties and near the instability. This is concluded
from former work \cite{su57,moden} and again in analogy to
the theory of static domains \cite{dom,domlin} and this is in
agreement to \cite{el88}. There the domain states are found
far above the Suhl instability. The wall profile, a local
property,
exhibits an oscillating structure compared to the step function
profile of this work.

Eq.(\ref{21}) describes undamped excitations, which result
from the high degeneracy
of the domain states. For realistic, infinite extended, models
this degeneracy is only partly removed. At least the
invariance due to  translations of domain walls remains,
leading again to undamped excitations. The presence of
undamped, or in finite systems, weakly damped excitations
can make it difficult to find experimentally  the corresponding
stationary state. The observed time
dependent effects \cite{ex} may, probably only within
certain regions, be related to these long living excitations.
These ideas certainly need further confirmation, but are
supported by the measurements of Wiese {\it et al.}\cite{ex}. At
low temperatures, where excitations are usually less important,
no time dependence was found in the power absorption.

I am indebted to H. Benner and G. Sauermann for discussions,
to H. J. Elmer for correspondence and for pointing out to me

ref.\cite{el88}. This work was performed within  SFB 185.


\begin{references}
\bibitem{com1} This success should not be overestimated,
                     as these phenomena are generic according
                     to the general concepts of non-linear
dynamics.
\end{references}
\end{document}